**Direct observation of magnetization dynamics generated by nano-contact spin-torque vortex oscillators**

P. S. Keatley[1]*, S. R. Sani[2,3,4], G. Hrkac[5], S. M. Mohseni[2,7], P. Dürrenfeld[7], T. H. J. Loughran[1], J. Åkerman[2,3,7], R. J. Hicken[1]

[1] *School of Physics and Astronomy, University of Exeter, Stocker Road, Exeter, EX4 4QL, UK.*
[2] *Materials and Nano Physics, School of ICT, KTH Royal Institute of Technology, Electrum 229, 164 60 Kista, Sweden.*
[3] *NanOsc AB, Electrum 205, 164 40 Kista, Sweden.*
[4] *Department of Physics & Astronomy, Uppsala University, Box 516, SE-751 20 Uppsala, Sweden.*
[5] *College of Engineering, Mathematics and Physical Science, University of Exeter, Exeter, EX4 4SB, UK.*
[6] *Department of Physics, Shahid Beheshti University, Tehran, Iran*
[7] *Physics Department, University of Gothenburg, Fysikgränd 3, 412 96 Gothenburg, Sweden*

Time-resolved scanning Kerr microscopy has been used to directly image the magnetization dynamics of nano-contact (NC) spin-torque vortex oscillators (STVOs) when phase-locked to an injected microwave (RF) current. The Kerr images reveal free layer magnetization dynamics that extend outside the NC footprint, where they cannot be detected electrically, but which are crucial to phase-lock STVOs that share common magnetic layers. For a single NC, dynamics were observed not only when the STVO frequency was fully locked to that of the RF current, but also for a partially locked state characterized by periodic changes in the core trajectory at the RF frequency. For a pair of NCs, images reveal the spatial character of dynamics that electrical measurements show to have enhanced amplitude and reduced linewidth. Insight gained from these images may improve understanding of the conditions required for mutual phase-locking of multiple STVOs, and hence enhanced microwave power emission.

Spin-torque oscillators (STOs) are nanoscale non-linear microwave devices with frequency that can be tuned by varying a DC bias current.[1-5] . The interaction of an STO with external microwave sources, [6-11] or other STOs, [12-19] is particularly rich and can promote understanding of other interacting non-linear oscillators ranging from semiconductor lasers [20] to neuromorphic circuits [21]. When multiple STOs are formed by depositing metallic nano-contacts (NCs) on a microscale spin valve mesa,[22-26] they may interact via the magnetization dynamics within their shared magnetic layers, leading to enhanced microwave power output and phase stability through mutual phase-locking.[12-15] The microwave power generated by auto-oscillations of a magnetic vortex[23] at MHz frequencies can be much larger than that generated by quasi-uniform precession at GHz frequencies.[4] Indeed enhancements of more than two orders of magnitude have been observed within the same device [15]. However, limited experimental progress has been made towards phase-locking multiple spin-torque *vortex* oscillators (STVOs) via the magnetization dynamics of their shared layers,[14, 19] and so improved understanding of the spatio-temporal character of the magnetization dynamics beyond the perimeter of an individual NC is now required.

Microwave emission occurs due to the giant magnetoresistance of the NC-STO[22] as the relative orientation of the spatially-averaged magnetizations of the layers varies in time in the region immediately beneath the NC. Therefore, magnetization dynamics outside of the NC perimeter cannot be detected by electrical measurements and instead are usually calculated numerically.[27] Very few experiments have been reported in which the magnetization dynamics outside a NC-STO have been directly mapped, and then only for STOs that exhibit precession rather than vortex gyration.[28-31] These experiments successfully mapped the spatial character of the precession, while only one revealed its time dependence.[31]

Here we report the direct experimental observation of magnetization dynamics in the vicinity of single and double NC-STVOs using time-resolved scanning Kerr microscopy. Stroboscopic Kerr images were acquired by phase-locking the STVOs to both an injected microwave (RF) current and the probing laser pulses. For both devices, magnetization dynamics were observed in the vicinity of the NCs when the vortex gyration frequency $f_G$ was locked (equal to) to that of the RF current $f_{RF}$. We refer to this as the primary locking range for which stable orbits of the vortex core are expected. The dynamics are complicated by the combined action of the spin-torque and the Oersted field $H_{RF}$ associated with the RF current. In the single NC device we show that these contributions can be separated by varying the DC current $I_{DC}$. Surprisingly, outside of the primary locking range, images of the single NC device reveal unreported partially-locked magnetization dynamics that can be obscured in electrical measurements by the residual injected RF current. In this case the vortex trajectory rotates at $f_{RF}$, while the vortex core traverses the trajectory at $f_G$. Characterization of the spatial area occupied by the dynamics reveals that the partially locked mode can extend further from the NC, suggesting that an injected RF current might be useful to initiate phase-locking of multiple STVOs.

In the double NC device, two collective modes of auto-oscillation were observed in the electrical measurements. The Kerr images reveal that these collective modes correspond to the coupled gyration of separate vortices associated with each NC, rather than a single vortex orbiting both NCs. Furthermore, the amplitude of microwave emission is largest when the vortices gyrate with similar amplitude and phase, as determined from time-resolved images of each mode separately injection locked at different values of $I_{DC}$. These measurements may provide insight into the conditions that favor mutual locking.

The NC-STVO devices were formed by depositing NCs onto spin-valve multilayer stacks that contained a Co reference layer and a NiFe free layer, Figure. 1(a).[32] The nominal diameter of the NCs in the single and double NCs devices was 250 nm and 100 nm respectively. When a sufficiently large value of $I_{DC}$ is applied to the NCs, the associated DC Oersted field $H_{DC}$ leads to the formation of a vortex anti-vortex pair within the otherwise uniform free layer magnetization.[27] Positive current is defined as electrons passing from the reference layer to the free layer. Therefore, for negative values of $I_{DC}$ the spin-torque acting on the vortex core can compensate the damping and excite gyrotropic auto-oscillations, which appear as peaks in the microwave emission spectrum, with frequencies $f_G$ (Figure. 1(b), (c)) that have a characteristic dependence upon $I_{DC}$.[15]

In TRSKM measurements, Fig. 1(a), three components of the free layer dynamic magnetization vector ($\Delta M_x$, $\Delta M_y$, $\Delta M_z$) were simultaneously detected using longitudinal and polar magneto-optical Kerr effects, and mapped with ~500 nm spatial resolution, and picosecond temporal resolution.[33, 34] The stroboscopic nature of TRSKM means that resonant modes can only be observed when the oscillation frequency is an integer multiple of the 80 MHz probe laser pulse repetition rate.[35] A small in-plane bias magnetic field $H_B$ (~3 to 5 mT) was applied to prevent switching of the free layer equilibrium magnetization. The free-running frequency of the selected STVOs, *i.e.* that in the absence of $I_{RF}$, was found to cross, or approach, 160 MHz as $I_{DC}$ was reduced, Figs. 1(b) and 1(c). These STVOs were therefore well suited for TRSKM using injection locking with an RF current of frequency $f_{RF}$ = 160 MHz and amplitude $I_{RF}$ = 1.4 mA. The RF current was also phase modulated (0 to 180°) at a frequency of ~3 kHz to enable phase sensitive detection of the dynamic Kerr signals. Kerr images in the primary locking range were acquired when the value of $I_{DC}$ was tuned such that $f_G = f_{RF} = 160$ MHz. For the single and double NC devices a single auto-oscillation and two collective auto-oscillations were observed (respectively Fig. 1(d) and (e)). As $|I_{DC}|$ is decreased, the mode frequencies exhibit a monotonic decrease before becoming locked to that of the injected RF current $f_{RF}$. When locked, $|I_{RF}| < 0.1|I_{DC}|$. Optical access to the spin valve free layer was achieved by careful design and fabrication of the microwave electrical contacts so that more than half of the area of the spin valve mesa remained uncovered while providing electrical contact to the NCs at the center of the mesa, Fig. 1(c).

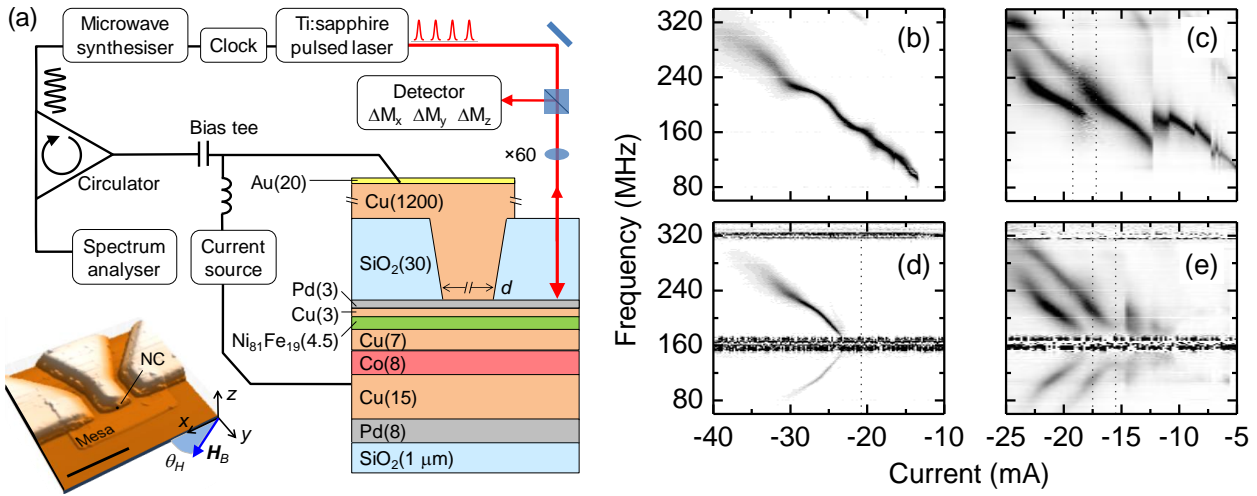

Figure 1. (a) A schematic of the combined apparatus for TRSKM and frequency-domain magneto-electronic measurements. Inset, an atomic force microscope image of the contact pad geometry allowing for optical access to the extended spin valve mesa (scale bar 10 μm). The approximate location of the NC (center of the mesa) is indicated by a small black dot. (b) and (c) The free-running signal generated by single and double NC devices subject to DC current only. White corresponds to 0 nV/Hz$^{1/2}$ and black corresponds to 4 nV/Hz$^{1/2}$ and 10 nV/Hz$^{1/2}$ in (b) and (c) respectively. Vertical dashed lines in (c) indicate the values of $I_{DC}$ (-19.2 mA and -17.2 mA) at which the spectra of Fig. 3(f) were extracted. (d) and (e) The signal generated by single and double NC devices subject to an additional RF current with frequency of 160 MHz and amplitude of 1.4 mA. White corresponds to 0 nV/Hz$^{1/2}$ and black corresponds to 2 nV/Hz$^{1/2}$ and 6 nV/Hz$^{1/2}$ in (d) and (e) respectively. In (d) and (e) the vertical dashed lines indicate the values of $I_{DC}$ (-20.8 mA in (d), and -17.5 mA and -15.5 mA in (e)) at which Kerr images were acquired. In (b) and (d) $|H_B| = 5$ mT and $\theta_B = 270°$. In (c) and (e) $|H_B| = 3$ mT and $\theta_B = 0°$.

In the single NC device two main regions of magnetization dynamics extending outside of the NC perimeter were generated and mapped as a function of time. Kerr images were acquired at 15° phase steps through one cycle of the RF current in the primary locking range at $I_{DC}$ = -20.8 mA (vertical dotted line, Fig. 1(d)) allowing movies of the magnetization dynamics to be compiled, see supplemental movie.[32] In Figure 2(a), large amplitude ($\Delta M_x \sim 0.5 M_s$) localized dynamics near to the end of the center contact pad (point 1 in the left-most panel of Fig. 2(a)) can be most clearly seen as white and black greyscale when the phase of the microwave current is 0° or 180° respectively. Corresponding dynamics in $\Delta M_y$ and $\Delta M_z$ can also be seen in Fig. 2(b) and (c) respectively. Since the RF current is phase modulated, the contrast represents the change in the magnetization between two dynamic states separated by 180°. As the phase of the microwave current is swept from 0° to 90° the area of the localized dynamics reduces in size and almost vanishes. At the same time the amplitude of the far-field dynamics (location 2, Fig. 2(a)) increases to a maximum at 90°, exhibiting white greyscale in $\Delta M_x$ that surrounds the localized dynamics and extends far into the upper half of the image. The diminution of the far field dynamics with increasing distance from the region of concentrated current, between the NC and the via to the ground plane, is expected from the approximate 1/$r$ dependence of the RF Oersted field.

The simulated images in Figure 2 were obtained by taking the difference of two simulated micromagnetic states separated in phase by 180°. While a comparison of the measured and simulated images reveals a strong similarity in the contrast of the observed localized dynamics at 180° (enclosed by dotted ellipses), the experiment shows that the dynamics extend further from the NC than expected from the simulations. Furthermore, the dynamics of the surrounding mesa are somewhat different and may, in part, be a consequence of different pinning sites for the anti-vortex in the experiment and the simulation. In the Kerr images we ascribe the oscillatory feature at the top-left corner of the contact pad to the anti-vortex (see arrows in 180° images), since it appears to be pinned approximately 2 μm from the NC. Previous reports have suggested such pinning of the anti-vortex far from the NC at the edge of a simulated mesh, as in our simulation (outside of region shown).[27] Here the anti-vortex could conceivably be pinned by stray electromagnetic fields from the contact pad.

The amplitude of the localized (region 1) and far-field (region 2) dynamics extracted from $\Delta M_x$ images reveal that the far-field dynamics lag the localized dynamics by 90°, Fig. 2(d). Furthermore, the far-field dynamics may instead be observed in $\Delta M_y$ images by rotating the small (~5 mT) bias magnetic field $H_B$ through 90°, c.f. Fig. 3(c). Therefore, we ascribe the dynamics of region 2 to excitation by the RF Oersted field (~0.1 mT), which exerts a torque on the quasi-uniform magnetization far from the NC.

In Fig. 2(d), the zero amplitude Kerr signal extracted from point 3 in Fig. 2(a) indicates that magnetization dynamics beneath the optically thick contact pads are not detected.

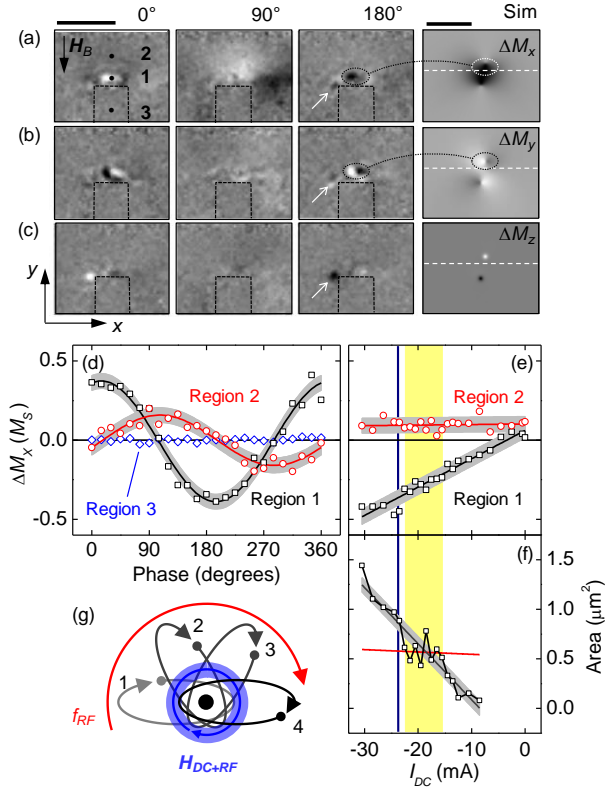

Figure 2. (a-c) Time- and vector-resolved Kerr images corresponding to the $\Delta M_x$ and $\Delta M_y$, (in-plane), and $\Delta M_z$ (out-of-plane) components of the dynamic magnetization (scale bar 5 μm) acquired at $|I_{DC}| = 20.8$ mA with $|H_B| = 5$ mT and $\theta_B = 270°$. The dotted black line depicts the approximate outline of the top contact pad. In (a and b) black/white represents a change in magnetization of +/- $0.5M_s$ where $M_s$ is the saturation magnetization. A micromagnetic simulation (Sim) shows the contrast expected for phase-modulated imaging (scale bar 200 nm, $|I_{DC}| = 18$ mA). (d) Time-resolved traces of $\Delta M_x$ corresponding to locations 1 (black symbols), 2 (red symbols), and 3 (blue symbols). (e) Amplitude of $\Delta M_x$ at location 1 (black symbols) and 2 (red symbols), and (f) area of the localized dynamics around location 1 extracted from $\Delta M_x$ images as a function of $I_{DC}$. The area was defined by a contour corresponding to $0.25M_s$ in $\Delta M_x$ images plotted on a contrast scale of 0 to $0.5M_s$. In (d-f) fitted curves are guides to the eye, while grey shading represents an uncertainty of one standard deviation derived from signals extracted from location 3. In (e and f) the vertical blue line represents the onset of locking, while the yellow band represents the primary locking range. (g) A simple schematic representation of the partially locked mode, which exhibits a change in the core trajectory at $f_{RF}$ due to the time-dependent amplitude of the Oersted field ($H_{DC+RF}$), while the core traverses the trajectory at $f_G$. The blue shading of $H_{DC+RF}$ illustrates the "breathing" of the azimuthal Oersted field lines. The center of the NC is represented by the black dot at the center of the diagram (not to scale).

If the localized dynamics were driven by the RF field, then they should be observed when $I_{DC} = 0$ mA. Instead their amplitude decreases monotonically as $|I_{DC}|$ is reduced, and vanishes when $|I_{DC}| < 5$ mA, Fig. 2(e). In contrast, the amplitude of the far-field dynamics remains constant since they are driven by the RF field, the amplitude of which remains constant as $|I_{DC}|$ is reduced. While this confirms that the localized dynamics are driven by spin-torque, their observation outside of the primary locking range (yellow band in Figs. 2(e and f)) suggests partial locking of the vortex dynamics to the RF current, which is not observed in the electrical measurements when $f_{RF} = 160$ Hz, Fig. 1(d). Evidence of partial locking in electrical measurements can instead be seen when $f_{RF} = 320$ MHz, see supplemental material [32], although the spatial character of the mode is then expected to be different. A possible explanation for the partially locked dynamics is that the core trajectory changes at $f_{RF}$, due to the time-dependent amplitude of the Oersted field ($H_{DC+RF}$) at $f_{RF}$ [10, 11], while the core traverses the trajectory at its free-running frequency $f_G$. The simplified schematic in Fig. 2(g) illustrates such a mode. In general the trajectory will not be circular, or concentric with the center of the NC (black dot, not to scale), and is therefore shown in Fig. 2(g) as a displaced ellipse, although it is likely to be more complicated in reality due to the precise equilibrium magnetic configuration of the free layer in the vicinity of the NC. Outside the locking range, at a particular phase of the RF current, the core will be at different locations on subsequent orbits (Fig. 2(g), 1-4). The trajectory is locked to the injected RF current, which is in turn phase modulated. Therefore, the detected Kerr signal corresponds to a change of the magnetization resulting from the transition between two trajectories separated by 180°, for example between ellipses 1 and 4. Since the gyration frequency is not locked, the magnetization within each trajectory is temporally averaged. Micromagnetic simulations support this interpretation, details of which are provided in the supplemental material.[32] The partially locked behavior allows the amplitude and area of the localized dynamics to be tuned more widely via the DC current, Figs. 2(e) and 2(f). Generally, a monotonic decrease in the area of localization was observed as $|I_{DC}|$ was reduced, while a plateau forms within the primary locking range (dotted red line). Here the vortex core and time-varying restoring potential are in phase, resulting in a more stable trajectory. In Fig. 2(e and f) the primary locking range (yellow band) is estimated from the width of the plateau and is in good agreement with the onset of locking (vertical blue line) observed in electrical measurements.

For the double NC device, the NCs had 100 nm diameter and 900 nm center-to-center separation. The NCs were electrically connected in parallel for nominally identical current injection, while individual control of core polarity of vortices nucleated at each NC was not possible. In the free-running electrical measurements of Fig. 1(c), a single mode is observed for -18.2 mA < $I_{DC}$ < -5 mA, before a second mode appears for $I_{DC} \leq$ -18.2 mA. Kerr images of these modes were acquired by injection locking at 160 MHz. In each case the value of $I_{DC}$ was chosen to lie within the locking range (vertical dotted lines in Fig. 1(e)), at -17.5 mA (Fig. 3(a)) and -15.5 mA (Fig. 3(b)) for the lower and higher frequency modes respectively.

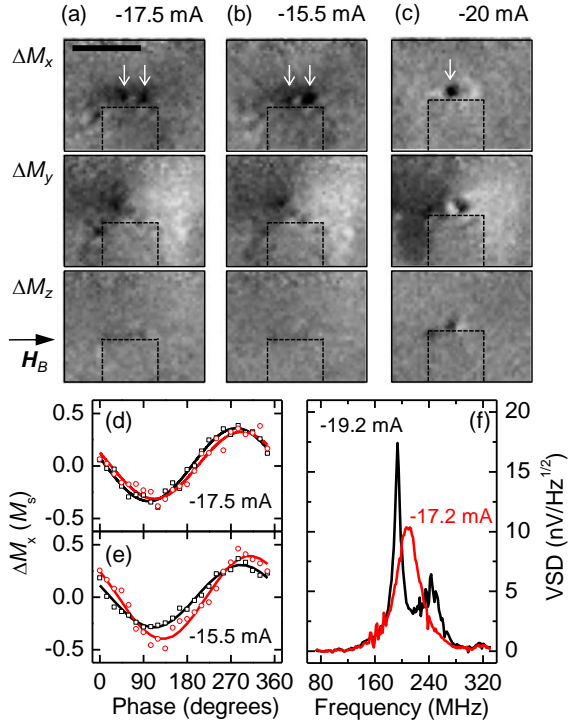

Figure 3 Spatial character of magnetization dynamics generated by a pair of STVOs. In (a) and (b) time- and vector-resolved Kerr images acquired at a fixed phase of the RF current at $I_{DC}$ = -17.5 mA and $I_{DC}$ = -15.5 mA show two regions of localized dynamics (indicated by arrows). In (a) and (b) $|H_B|$ = 3 mT and $\theta_B$ = 0°. Black/white greyscale represents +/ 0.4$M_s$ respectively. (c) The equivalent response for a single NC shows only a single region of localized dynamics (arrow), where black/white represents +/-0.5$M_s$. In (c) $|H_B|$ = 5 mT and $\theta_B$ = 0°. (d) and (e) Time-resolved traces extracted from the left (black symbols) and right (red symbols) region of localized dynamics, indicated by the arrows in the $\Delta M_x$ images, acquired over one RF cycle for $I_{DC}$ values of -17.5 mA (d) and -15.5 mA (e). Fitted curves with frequency 160 MHz are guides to the eye. (f) Free-running spectra extracted along the vertical dashed lines in Fig. 1(c) that lie to either side of the mode transition occurring at $I_{DC}$ = -18.2 mA.

In contrast to the single NC (Fig. 3(c)), the double NC device exhibits two regions of localized dynamics confirming that a separate vortex is formed at each NC, Figs. 3(a) and 3(b) (arrows), and supplemental movies.[32] When $I_{DC}$ = -17.5 mA, these regions exhibit similar amplitude and phase (Fig. 3(d)). However, for $I_{DC}$ = -15.5 mA the right NC exhibits greater amplitude and area of localization with respect to the left NC, and also different phase (Fig. 3(e)), which is expected to lead to reduced microwave emission. Spectra from the free-running response, obtained at $I_{DC}$ values within the one and two mode regimes (vertical dotted lines, Fig. 1(c)) are shown in Fig 3(f). The two modes observed for $I_{DC}$= -19.2 mA have reduced linewidth compared to the single mode at $I_{DC}$ = -17.2 mA, while the lower frequency mode at $I_{DC}$= -19.2 mA also has enhanced amplitude. This suggests that for small values of $|I_{DC}|$ the radii of gyration are sufficiently small that the vortices gyrate almost independently. The different response of each NC observed in the $\Delta M_x$ images of Fig. 3(b) may therefore be ascribed to nanoscale structural differences of the NCs, which in turn affect the current distribution and Oersted field spatial character. The resulting lack of mutual coherence leads to the broad single mode in the free-running spectrum (Fig. 3(f), red curve). However, when the value of $|I_{DC}|$ and the radii of gyration increase, the vortices interact more strongly. An in-phase mode is observed in the $\Delta M_x$ images (Fig. 3(a)) and the enhanced mutual coherence leads to collective modes of reduced linewidth in the free-running electrical spectra (Fig 3(f), black curve). To obtain a more quantitative description, improved models are required that describe injection and mutual locking on an equal footing. However, the mode character observed in the injection-locked images is consistent with that observed in the free-running electrical measurements and so may provide a means by which to understand the conditions required for optimum microwave emission in a mutually-locked state.

In summary, TRSKM has been used to characterize the injection-locked magnetization dynamics that extend outside the perimeter of NC-STVOs. Fully-locked states, and partially-locked states of increased spatial extent, were detected for the single NC device. An in-phase collective mode was observed in a double NC device that was consistent with the enhanced amplitude and reduced linewidth observed in the free-running microwave emission spectrum. Injection-locked TRSKM measurements may therefore provide new insight into the conditions required for mutual phase-locking of STVOs.

The authors gratefully acknowledge the financial support of the Engineering and Physical Sciences Research Council under grants EP/I038470/1 and EP/K008501/1, the Royal Society under grant UF080837, the Swedish Research Council (VR), the Swedish Foundation for Strategic Research (SSF), and the Knut and Alice Wallenberg Foundation (KAW).